\documentclass[%
 aip, pop,
 amsmath,amssymb,
reprint,
]{revtex4-1}
\usepackage{graphicx}
\usepackage{dcolumn}
\usepackage{bm}
\usepackage[utf8]{inputenc}
\usepackage{soul}
\usepackage[T1]{fontenc}
\usepackage{amsmath}
\usepackage{gensymb}
\usepackage{tikz}
\usetikzlibrary{shapes.geometric, positioning}
\usetikzlibrary{backgrounds}
\usetikzlibrary{calc}
\usepackage{soul}
\setstcolor{blue}
\usepackage{subcaption}
\usepackage{natbib}
\usepackage{xparse}

\newcommand{\ab}[1]{{#1}}
\newcommand{\av}[1]{{#1}}

\newcommand{\fsa}[1]{\left\langle #1 \right\rangle}
\newcommand{\vv}[1]{\mathbf{#1}}

\NewDocumentCommand{\pdv}{m g}{
  \IfNoValueTF{#2}
    {\frac{\partial}{\partial #1}}
    {\frac{\partial #1}{\partial #2}}
}

\begin{document}

\title{Update on the design of the Columbia Stellarator eXperiment}

\author{A. Baillod}
\email[]{ab5667@columbia.edu}
\affiliation{Department of Applied Physics and Applied Mathematics, Columbia University, New York, New York 10027, USA}
\author{A. Veksler}
\affiliation{Department of Applied Physics and Applied Mathematics, Columbia University, New York, New York 10027, USA}
\author{R. Lopez}
\affiliation{Department of Applied Physics and Applied Mathematics, Columbia University, New York, New York 10027, USA}
\author{D. Schmeling}
\affiliation{Department of Applied Physics and Applied Mathematics, Columbia University, New York, New York 10027, USA}
\author{M. Campagna}
\affiliation{Department of Applied Physics and Applied Mathematics, Columbia University, New York, New York 10027, USA}
\author{E. J. Paul}
\affiliation{Department of Applied Physics and Applied Mathematics, Columbia University, New York, New York 10027, USA}
\author{A. Knyazev}
\affiliation{Department of Applied Physics and Applied Mathematics, Columbia University, New York, New York 10027, USA}
\date{\today}

\begin{abstract}
We present the final configuration chosen to be build for the Columbia Stellarator eXperiment (CSX), a new stellartor experiment at Columbia University. In a recent publication, Baillod \textit{et.al.} (NF, 2025) discussed in detail the different objectives, constraints, and optimization algorithms used to find an optimal configuration for CSX. In this paper, we build upon this first publication and find a configuration that satisfies all the constraints. We describe this final configuration including discussion of the coil finite build effects, sensitivity analyses, and the plasma neoclassical physics properties using the SFINCS code. These post-processing calculations provide \av{\st{a}} confirmation that the experimental goals of CSX can be achieved with the presented configuration.
\end{abstract}

\maketitle 

\section{Introduction}
The Columbia Stellarator eXperiment (CSX) is a new university-scale stellarator currently under construction at Columbia University. The project is motivated by three complementary objectives that span the domains of plasma physics, engineering innovation, and education. From a physics standpoint, CSX will provide a versatile platform to investigate neoclassical physics in magnetic configurations close to quasi-axisymmetry (QA). From an engineering perspective, the device seeks to demonstrate the feasibility of employing non-insulated, high-temperature superconducting (NI-HTS) coils in a stellarator geometry. In parallel, CSX fulfills an educational mission by offering hands-on training opportunities for students and researchers in stellarator design, construction, and operation, thereby supporting the development of the next generation of fusion scientists and engineers.

The CSX device consists of two planar, water-cooled copper coils --- referred to as poloidal field (PF) coils --- and two interlinked (IL) coils fabricated with NI-HTS tape. \ab{One limiting factor of HTS tape magnets is the mechanical stresses and strains that can be sustained by HTS magnets. This impacts the coil shape and complexity that can be achieved, and needs to be taken into account when optimizing the coils.} The IL coils are housed within a cylindrical vacuum vessel, while the PF coils are mounted outside the vessel. Both the vacuum vessel, and the PF coils, are repurposed from the Columbia Non-neutral Torus (CNT) experiment.\citep{pedersen_2004b,pedersen_2006a} This compact design uses the same coil topology as for the CNT experiment and other proposed designs, such as the Compact Stellarator with Simple Coils (CSSC).\citep{yu_2022,qiu2025optimizationcompactstellaratorsimple}

In previous work, \citet{baillod_integrating_2025} presented the optimization framework that guided the initial design of CSX. That study integrated both magnetic and engineering constraints, including limits on coil length, HTS strain, and available winding space, and produced a set of five candidate configurations that satisfied the principal design objectives. Building on that foundation, the present work refines the optimization procedure to incorporate new constraints identified during the coil prototyping phase. In particular, we introduce a finite-build model of the NI-HTS coils, based on an adaptation of the multi-filament representation developed by \citet{singh_2020}, to more accurately model the magnetic field and capture HTS strain across the different filaments.

Using this improved framework, we perform a comparative assessment of the candidate configurations in terms of their magnetic performance, engineering feasibility, and sensitivity to manufacturing and alignment errors. Based on this analysis, a single configuration is selected for construction. The following sections present a detailed description of the final CSX design, summarize its key metrics, and assess its physics performance through estimates of neoclassical flow damping timescales obtained using the drift-kinetic code SFINCS.~\citep{landreman_2014}

\section{An improved optimization scheme}\label{sec.optimization}

In \citet{baillod_integrating_2025}, the different physics objectives and engineering constraints of CSX were discussed in detail. We list these below for completeness and refer the reader to our earlier publication for a full discussion on this topic. The main objective of CSX is to reach a QS error below $8$\%, in a plasma volume larger or equal to $0.1\ \text{m}^3$, with a rotational transform of the order of $0.27$. Furthermore, the IL coils must fit within the existing vacuum vessel and sufficient space should be left between both coils and between the coils and the plasma surface. The HTS strain has to remain below $0.22\%$, the coil length below $5$m, and it must be possible to wind the coils without significant difficulties.

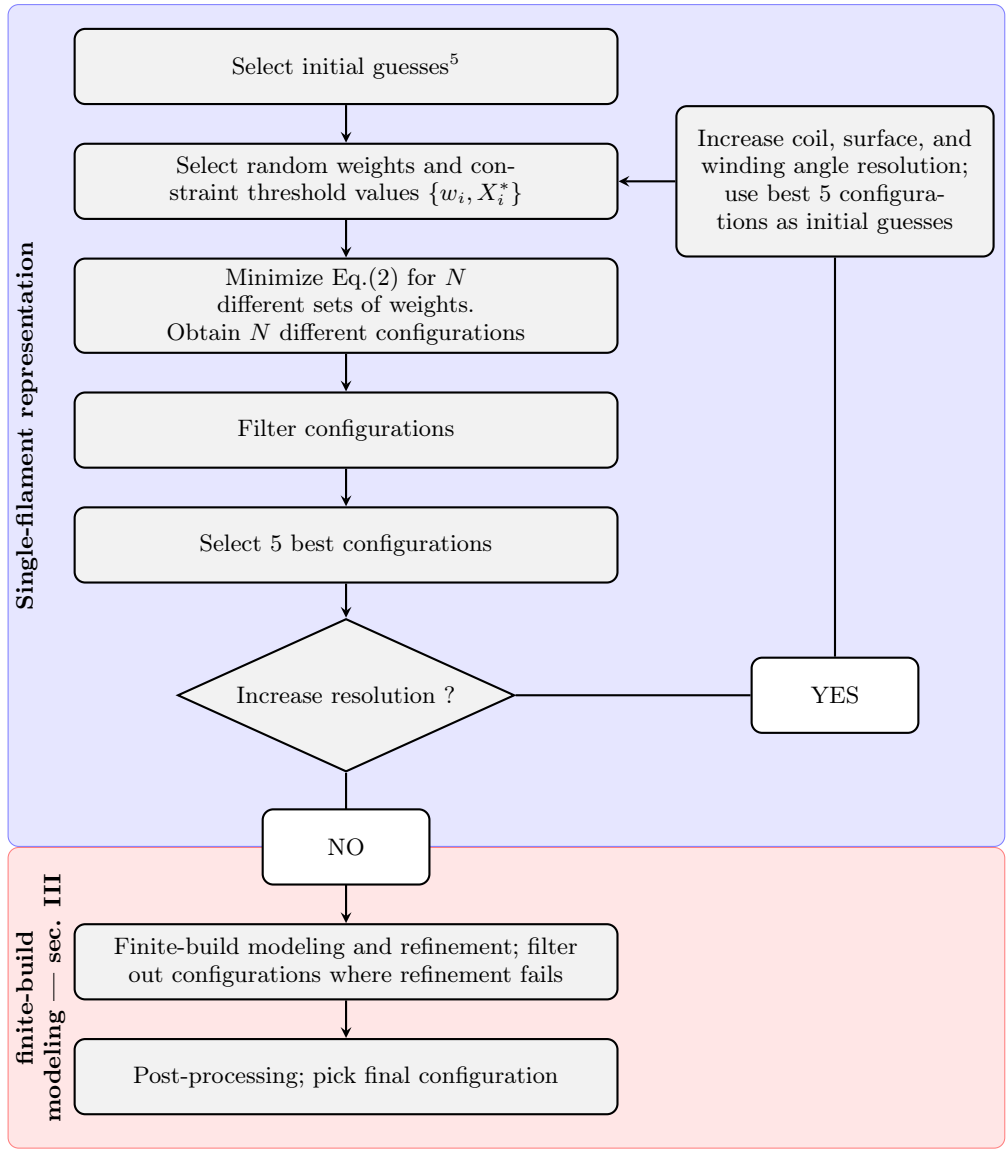
\begin{figure*}
    \centering
    \begin{tikzpicture}[
        node distance=2cm,
        background/.style={rounded corners},
        box/.style={
            rectangle,
            rounded corners,
            draw=black,
            thick,
            fill=gray!10,
            text width=7cm,
            align=center,
            minimum height=1cm
        },
        narrowbox/.style={
            rectangle,
            rounded corners,
            draw=black,
            thick,
            fill=gray!10,
            text width=4cm,
            align=center,
            minimum height=2cm
        },
        smallbox/.style={
            rectangle,
            rounded corners,
            draw=black,
            thick,
            fill=white,
            text width=2.0cm,
            align=center,
            minimum height=1cm
        },
        decision/.style={
            diamond,
            aspect=2.2,
            draw=black,
            thick,
            fill=gray!10,
            text width=4cm,
            align=center,
            inner sep=0pt
        },
        arrow/.style={
            -stealth,
            thick
        }
    ]

    \node[box] (step1) {Select initial guesses \citep{baillod_integrating_2025}};
    \node[box, below=.5cm of step1] (step2) {Select random weights and constraint threshold values $\{w_i,X_i^*\}$};
    \node[box, below=.5cm of step2] (step3) {Minimize Eq.(\ref{eq.objective boozer}) for $N$ \\different sets of weights. \\Obtain $N$ different configurations};
    \node[box, below=.5cm of step3] (step4) {Filter configurations};
    \node[box, below=.5cm of step4] (step5) {Select 5 best configurations};
    \node[decision, below of=step5] (step6) {Increase resolution ?};
    \node[smallbox] (tmp0) at ($(step6) + (6.5,0)$) {YES}; 
    \node[narrowbox] (tmp1) at ($(step2) + (6.5,0)$) {Increase coil, surface, and winding angle resolution; use best 5 configurations as initial guesses};
    \node[box, below=2cm of step6] (step7) {Finite-build modeling and refinement; filter out configurations where refinement fails};
    \node[box, below=.5cm of step7] (step8) {Post-processing; pick final configuration};

    \draw[arrow] (step1) -- (step2);
    \draw[arrow] (step2) -- (step3);
    \draw[arrow] (step3) -- (step4);
    \draw[arrow] (step4) -- (step5);
    \draw[arrow] (step5) -- (step6);
    \draw[arrow] (step6) -- (tmp0) -- (tmp1) -- (step2);
    \draw[arrow] (step6) -- (step7);
    \draw[arrow] (step7) -- (step8);

    \path (step6) -- (step7)
        node[smallbox, midway, above, yshift=-.5cm] (text_node_1) {NO};

    \begin{scope}[on background layer]
        \node[background, fill=blue!10, draw=blue!50,  minimum width=13.25 cm, minimum height=11.2 cm, anchor=south west] 
            (bg1) at ($(-4.5,0)!(text_node_1)!(-4.5,1)$) {};
        \node[background, fill=red!10, draw=red!50,  minimum width=13.25 cm, minimum height=4 cm, anchor=north west] 
            (bg2) at ($(-4.5,0)!(text_node_1)!(-4.5,1)$) {};
    \end{scope}
    
    \node[rotate=90, anchor=north, align=center] at (bg1.west) {\textbf{Single-filament representation}};
    \node[rotate=90, anchor=north, align=center] at (bg2.west) {\textbf{finite-build} \\ \textbf{modeling --- sec. \ref{sec:finit_build}}};

    \end{tikzpicture}
    \caption{CSX design scheme. The first section, highlighted in blue, is performed with a single-filament representation of the coils. The second section, in red, uses a multi-filament representation of the coils to better model the magnetic field generated by the coils.}
    \label{fig:optimization_scheme}
\end{figure*}

In \citet{baillod_integrating_2025}, the Boozer-surface approach \citep{giuliani_2022a,giuliani_2022} was shown to produce CSX configurations with lower quasi-axisymmetry (QA) error than alternative optimization methods when the relevant engineering constraints were included. For this reason, all results in the present study rely on this approach.

The Boozer-surface algorithm optimizes a given coil set $\mathcal{C}$. At each iteration, the magnetic field $\mathbf{B}$ produced by the coils is evaluated throughout space using the Biot–Savart law. An approximate magnetic surface $\Gamma$ is then constructed by solving a partial differential equation, denoted schematically as $\mathbf{r}(\mathcal{C})=0$ and described in detail in \citet{giuliani_2022,giuliani_2022a}. Various physics objectives --- such as rotational transform, QA error, enclosed volume, and aspect ratio — are evaluated on $\Gamma$. Derivatives of these objectives with respect to coil degrees of freedom are obtained using an adjoint method.
The CSX optimization objective function introduced in \citet{baillod_integrating_2025} can be written as
\begin{align}
f^\text{Boozer}(\mathcal{C}) &= \frac{1}{2} \iint |\mathbf{r}|^2 dS
+ \sum_i w_i, f_{i}^\text{plasma}(\Gamma(\mathcal{C}))  \\
&+ \sum_i w_i, f_{i}^\text{reg}(\mathcal{C}), \nonumber
\label{eq.objective boozer}
\end{align}
where the first term penalizes deviations of $\Gamma$ from a true magnetic surface, thereby coupling the plasma geometry to the coils. The second term contains physics objectives $f_i^\text{plasma}$ (e.g., rotational transform or QS error), and the third term contains engineering constraints $f_i^\text{reg}$ (e.g., HTS strain, coil–coil spacing, and coil–plasma distance). The weights $w_i$ specify the relative importance of each contribution.

In the optimization framework described in \citet{baillod_integrating_2025}, engineering constraints were formulated alongside physics objectives. Since then, coil prototyping has revealed additional challenges for NI-HTS winding. In particular, concave regions of the tape were found difficult to wind under tension, especially in the presence of significant torsion.  

To address this, we introduce a new penalty term,  
\begin{multline}
J_\text{concavity} = \oint \min\!\left(\hat{\mathbf{n}}_\text{frenet}(l)\cdot\hat{\mathbf{n}}_\text{HTS}(l),0\right)^2 \, \\ \times \max(\tau(l)-\tau^*,0)^2 \, dl,
\end{multline}  
where the integral runs along the coil centerline, $\hat{\mathbf{n}}_\text{frenet}$ is the Frenet--Serret normal, $\hat{\mathbf{n}}_\text{HTS}$ is the tape normal, $\tau(l)$ is the tape torsion, and $\tau^*$ is a tolerable threshold. This penalty is nonzero only in coil segments that are simultaneously concave and subject to torsion above $\tau^*$.  

In addition to incorporating $J_\text{concavity}$, the objective function has been refined to reflect updated engineering thresholds --- such as coil-to-vessel spacing --- that have become more precisely defined during the ongoing coil design process. These updated engineering constraints explain why the configurations presented in \citet{baillod_integrating_2025} were not considered for CSX. Instead, new configurations were sought.

One crucial objective that was missing in our earlier optimizations was the magnetic well, which is a proxy to providing stability against interchange modes.\citep{mercier1962critere,mercier1964equilibrium} We noticed that the magnetic well competes with the QA quality, and both could note be obtained simultaneously in CSX. It was decided that the QA objective was more important for the planned experiments of CSX, and decided to drop the objective on the magnetic well in what follows.

Most engineering constraints are implemented as quadratic penalties of the form
\begin{equation}
f_i^\text{reg}(\mathcal{C}) = \max[\pm (X(\mathcal{C}) - X_i^*), 0]^2,
\end{equation}
where $X(\mathcal{C})$ is a coil-dependent metric and $X_i^*$ is its threshold value. The sets $\{w_i\}$ and $\{X_i^*\}$ constitute hyperparameters of the optimization scheme, and their values strongly influence the configurations to which the optimizer converges.

Since our earlier publication, the optimization scheme of CSX evolved to automatically choose hyper-parameters. The new optimization scheme is summarized in Figure \ref{fig:optimization_scheme}. The first step of the optimization consists on picking suitable initial guesses. Here we use a rescaled version of the compact stellarator with simple coils (CSSC)\citep{yu_2022} named \emph{rCSSC}, an optimized QA stellarator proposed by \citet{jorge_2023} named \emph{jorge\_QA}, or one of the CNT configurations, here named \emph{CNT32}. Details about these initial guesses are provided in \citet{baillod_integrating_2025}. 

The next steps are designed to automatically pick the optimization hyperparameters. To avoid manually tuning these hyperparameters, they are instead sampled randomly within physically reasonable bounds. The optimization is then repeated for many different hyperparameter samples. Because the minimization of Eq.~\ref{eq.objective boozer} is a non-convex problem with numerous local minima, each hyperparameter set slightly reshapes the objective landscape and generally leads to a different optimum.

Configurations that violate engineering constraints are discarded. Among the remaining solutions, the five closest to quasisymmetry are retained. For low-resolution runs, these five configurations are used as initial guesses for subsequent optimizations at higher resolution. Once the highest resolution is reached, the best configurations are selected for detailed analysis. The overall procedure is summarized in Fig.~\ref{fig:optimization_scheme}. This staged workflow eliminates the need for manual weight tuning: only reasonable bounds on the hyperparameters must be chosen. By exploring a family of modified objective landscapes, the method helps the optimizer escape unfavorable regions of parameter space.

In practice, we use four optimization stages with boundary resolutions $(M,N)=\{(6,6), (8,8),(10,10), (12,12)\}$ and coil resolution $N_c=\{5,7,8,9\}$. The resolution of the winding angle is set to $N_\alpha=10$. At each stage, approximately $500$ optimizations are performed with randomized hyperparameters. In the first stage, three distinct initial configurations are used --- \emph{CNT32}, \emph{rCSSC}, and \emph{jorge\_QA} --- as described in \citet{baillod_integrating_2025}. The optimization is performed using \emph{simsopt},\citep{landreman_2021} a Python framework for stellarator design, together with the BFGS algorithm \citep{liu_1989} implemented in \emph{scipy.optimize}.\citep{Virtanen2020}

As an outcome of this first optimization loop, we find high resolution configurations that satisfy the engineering constraints. These configurations were obtained with a filamentary coil representation; in what follows, we will refer to them as the \emph{single-filament configurations} (SFC), opposite to the \emph{multi-filament configurations} (MFC), that we discuss in the following section.

\section{Finite-build modeling}\label{sec:finit_build}
\av{The optimization loop described above and shown in the blue area of Fig.~\ref{fig:optimization_scheme} considers coils of infinitesimal width, hereafter referred to as filamentary coils.} However, real physical coils have a finite size, and the magnetic field produced by finite size coils is not the same as the one generated by filamentary coils. \av{The goal of finite-build modeling is to characterize the impact of the finite-build coils on CSX equilibria, and, if necessary, re-optimize the equilibria to create MFC's to minimize any deleterious effects.}

\ab{Here the modeling of the impact of finite-build coils follows the approach in \citet{singh_2020}. The typical single-filament representation is extended to include multiple filaments, which are distributed across the volume of the coil. Each filament carries an equal fraction of the total coil current.} \av{By evenly distributing the current from a single filament across multiple filaments, we create more realistic distribution of current. The magnetic field produced from each filament is computed using the Biot-Savart law.  Finally, the total magnetic field is computed by summing the magnetic field produced from each filament.}

We start with the filamentary coil \ab{representation used in the} single stage optimization, with position $\mathbf{r}(l)$. Associated with the \ab{coil geometry, we define the centroid frame $\lbrace \hat{\mathbf{t}}(l), \hat{\mathbf{n}}(l), \hat{\mathbf{k}}(l) \rbrace$, as defined in \citet{singh_2020}. This choice of frame removes singularities present in the classic Frenet-Serret frame, where a locally straight coil has an undefined normal vector, and near-straight coils may have large oscillations in the orientation of normal and bi-normal vectors over a short distance.}

Rotation of the HTS tape stack around the central, filamentary coil is used as an optimization free parameter to reduce coil strain. The frame of the HTS tape stack is defined by rotating the centroid frame normal and binormal vectors  $\hat{\mathbf{n}}(l)$ and $\hat{\mathbf{k}}(l)$ about $\hat{\mathbf{t}}(l)$ by an angle $\alpha(l)$.

\begin{align}
    \hat{\mathbf{n}}_{\text{HTS}}(l)=\hat{\mathbf{n}} \cos(\alpha(l))-\hat{\mathbf{k}} \sin(\alpha(l))\\ 
    \hat{\mathbf{k}}_{\text{HTS}}(l)=\hat{\mathbf{n}} \sin(\alpha(l))+\hat{\mathbf{k}} \cos(\alpha(l)). 
\end{align}
\ab{The HTS tapes are then stacked in the $\mathbf{n}_\text{HTS}$ direction, as sketched in Figure \ref{fig:coil_cross_section}.}

In the multifilament representation, the location of the original single filament representation, $\mathbf{x}_{\text{sf}}(l)$, becomes \av{the centre of the HTS tape stack cross section}. The positions of the filaments used to model the finite-build coil are defined as 
\begin{equation}
    \mathbf{x}_{\text{fb}}(l)=\mathbf{x}_{\text{sf}}(l) +\Delta n \ \mathbf{\hat{n}}_{\text{HTS}}(l)+\Delta k \ \mathbf{\hat{k}}_{\text{HTS}}(l),
    \label{eq.pos_mf}
\end{equation}
where $\mathbf{\hat{n}}_{\text{HTS}}(l)$ and $\mathbf{\hat{k}}_{\text{HTS}}(l)$ are the normal and binormal unit vectors of the HTS tape stack frame, and $\Delta n$ and $\Delta k$ are constants \ab{specific to each filament}.

The CSX coils are designed with four parallel HTS channels (see Figure \ref{fig:coil_cross_section}), each holding 250 winds of HTS tape. Previous design were based on a double channel design; a four channel design was deemed more suitable as it reduces the channel depth, which directly eases the winding process, and it reduces the field error generated by finite build effects. The four channel coils are also more compact than their two channel counterpart, which helps satisfying geometrical constraints.


\subsection{Finite-build effects on magnetic field error}\label{sec:fb_modeling_B}
\citet{mcgreivy_optimized_2021} note that the impact of finite size coils on magnetic field error scales as $O((\delta/L)^2)$, where $\delta$ is the coil \av{cross-section} size and $L$ is the coil to plasma distance. \av{For CSX, we find $(\delta/L)^2\approx (0.03 \text{ m}/0.1 \text{ m})^2=0.09$, which is non-negligible. This figure of merit is comparable to HSX\cite{almagri_1998-vbx55, singh_2020}, NCSX\cite{pomphrey_innovations_2001, williamson_modular_2005}, and W7-X\cite{Carlton-Jones_Paul_Dorland_2021, bonifetto_modeling_2011}, which have $(\delta/L)^2$ $\approx$ 0.5, 0.35, and 0.27 respectively. }

\begin{figure}
    \centering
    \includegraphics[width=\linewidth]{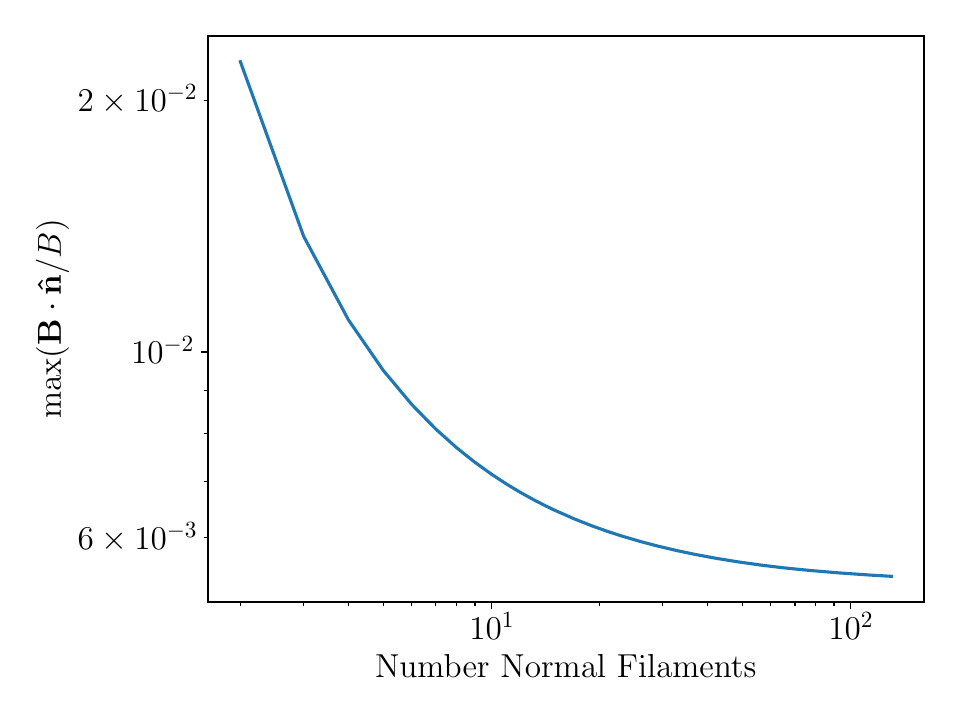}
    \caption{Scaling of normalized magnetic field error with increasing number of normal filaments.}
    \label{fig:filament_number_scaling}
\end{figure}

\av{To reduce the magnetic field error produced by finite-build coils, we optimize directly the finite-build configuration.} It is computationally favorable to use \ab{a small} number of filaments, as this decreases the number of Biot-Savart evaluations required in each step of optimization. The number of filaments to accurately determine the finite-build field is determined via a numerical convergence test of when the field error evaluated on the plasma boundary saturates with respect to increasing filament resolution. \av{Finite-build modeling primarily used 8 normal filaments and 1 binormal filament as a first pass refinement of good SFC into MFC. A higher resolution run with 80 normal filaments was performed on the SFC that produced the best MFC. At this resolution, the field calculation is converged, as shown in Fig.~\ref{fig:filament_number_scaling}. We confirmed that the relevant plasma metrics and profiles did not change between lower and higher resolution MFC's.}

\subsection{Finite-build effects on strain across tape stack}

Another metric examined in the context of finite-build coils is the mechanical strain on the \ab{HTS tape}. The strain has a functional dependence on the geometric curvature of the tape, which changes across the tape stack compared to the curvature of a filamentary coil. \ab{To compute the variation of strains in a finite-build coil, we extend below the methodology of \citet{paz-soldan_non-planar_2020}}. 

\begin{figure}
    \centering
    \includegraphics[width=\linewidth]{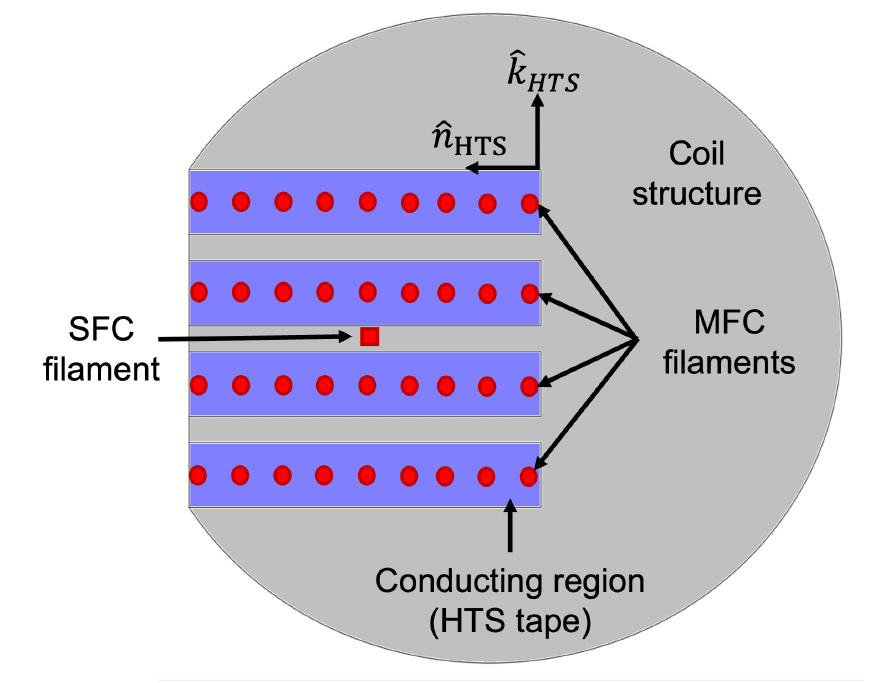}
    \caption{Visualization of filament placement for strain for MFC's. The filaments are placed in the center of the HTS tape stack in the $\mathbf{\hat{k}}_{\text{HTS}}$ direction, as required in \citet{paz-soldan_non-planar_2020}.}
    \label{fig:coil_cross_section}
\end{figure}

\ab{While multiple filaments per wind of HTS tape could be used to model the finite-build magnetic field, it is important to use only one filament per wind HTS tape for strain modeling, as what matters is the geometry of the line central to the HTS tape. Therefore, the position of the filaments may not be the same when evaluating the magnetic field versus the HTS tape strain. In what follows, we write $\mathbf{x}_{\text{fb}}$ to represent the position of a finite-build filament at the center (in the binormal direction) of the HTS tape stack, \cite{paz-soldan_non-planar_2020} as illustrated in Figure~\ref{fig:coil_cross_section}.}

A critical point is that one cannot use the frame of the single filament representation to describe the orientation of the finite-build filaments. The position vector of a filament is given by Eq.~\ref{eq.pos_mf}. Differentiating Eq.~\ref{eq.pos_mf} to compute the tangent vector $\mathbf{\hat{t}}_{\text{fb}}$ immediately shows that the tangent vector is \ab{different from the single filament tangent vector $\mathbf{t}(l)$}. Thus, the set $\lbrace \mathbf{\hat{t}}_{\text{fb}}, \mathbf{\hat{n}}_{\text{HTS}}, \mathbf{\hat{k}}_{\text{HTS}} \rbrace$ does not form an orthonormal basis, and cannot be used to describe the orientation of the finite-build filaments.

\begin{figure}
    \centering
    \includegraphics[width=\linewidth]{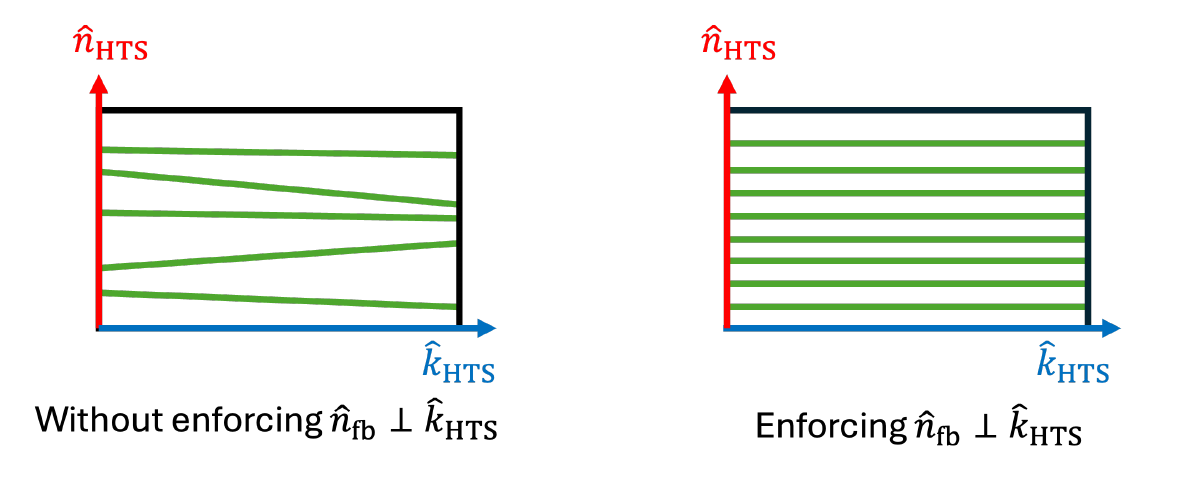}
    \caption{Comparison of HTS tape stack cross-sections with and without enforcing that $\mathbf{\hat{n}}_{\text{fb}} \perp \mathbf{\hat{k}}_{\text{HTS}}$.}
    \label{fig:finite-build-tape-stack-motivation}
\end{figure}

To determine the correct frame to describe the orientation of the finite-build filaments, we turn towards engineering constraints. The channels that hold the HTS tape stacks are constructed using the position of the single filament curve and an extrusion in the direction of $\mathbf{\hat{n}}_{\text{HTS}}$. Thus, the tape stack channel is rectangular in the plane defined by $\mathbf{\hat{n}}_{\text{HTS}}$ and $\mathbf{\hat{k}}_{\text{HTS}}$. For the individual tapes to fully fit in the channel, we require that $\mathbf{\hat{n}}_{\text{fb}} \perp \mathbf{\hat{k}}_{\text{HTS}}$. The failure of this requirement is \ab{illustrated} in Figure~\ref{fig:finite-build-tape-stack-motivation}.

This constraint allows us to define a local frame for the finite-build filaments. We construct it via a Gram-Schmidt process between $\mathbf{\hat{k}}_\text{HTS}$ and $\mathbf{\hat{t}}_{\text{fb}}$, which subtracts out the part of $\mathbf{\hat{k}}_\text{HTS}$ that is non-orthogonal to $\mathbf{\hat{t}}_{\text{fb}}$.
\begin{align}
    \mathbf{\hat{t}}_{{fb}}&=\frac{\mathbf{x}_{{fb}}'}{||\mathbf{x}'_{{fb}}||}\label{eq.t_mf} \\
    \mathbf{k}_{{fb}} &= \mathbf{\hat{k}}_\text{HTS} - (\mathbf{\hat{k}}_\text{HTS}\cdot \mathbf{\hat{t}}_{{fb}})\mathbf{\hat{t}}_{{fb}} \\
    \mathbf{\hat{k}}_{{fb}}&= \frac{\mathbf{k}_{{fb}}}{||\mathbf{k}_{{fb}}||} \\
    \mathbf{\hat{n}}_{{fb}} &= \mathbf{\hat{k}}_{{fb}} \times \mathbf{\hat{t}}_{{fb}}.
    \label{eq.alpha_frame}
\end{align}


The strain is computed from the change in the frame orientation over the length of the coil. The torsion and binormal curvature are given by\citep{Huslage_Paul_Haque_Gil_Foppiani_Smoniewski_Stenson_2025}
\begin{align}
\tau = \frac{\mathbf{\hat{k}}_{fb}\cdot \mathbf{\hat{t}}'_{fb}}{||\mathbf{x}'_{\text{fb}}||}
\label{eq.tor_curvature}
\\
\eta = \frac{\mathbf{\hat{k}}_{fb}\cdot \mathbf{\hat{n}}'_{fb}}{||\mathbf{x}'_{\text{fb}}||}. 
\label{eq.binorm_curvature}
\end{align}

The torsional strain and binormal curvature strain are then evaluated as
\begin{align}
\epsilon_{\tau} &= \frac{\tau^2w^2}{12} \\
\epsilon_{\eta} &= \frac{\eta w}{2},
\label{eq.strain}
\end{align}
where $w$ is the HTS tape width.


\subsection{Configuration refinement with finite-build modeling}

\begin{figure}
    \centering
    \includegraphics[width=\linewidth]{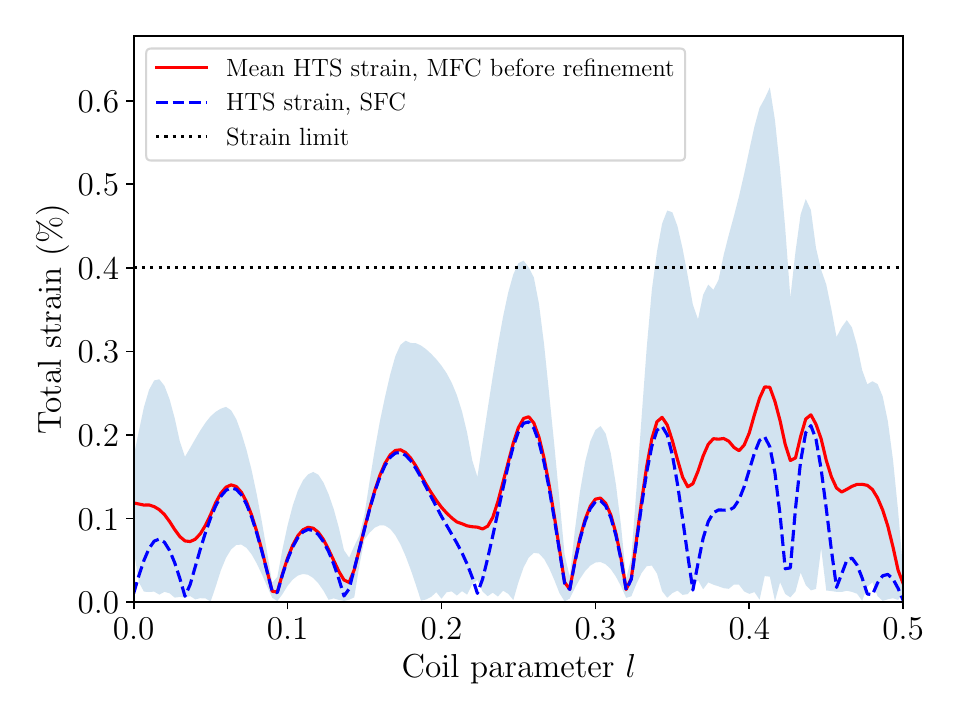}
    \caption{Strain for a CSX coil compared between a SFC and its MFC counterpart, before re-optimizing the MFC to reduce strain. The shaded area represents the variation between the minimum and maximum strain across the HTS tape stack. The maximum strain in certain filaments exceeds the strain threshold of the HTS tape.}
    \label{fig:finite-build-strain}
\end{figure}

\av{When converting from filamentary coils to finite-build coils, there is an increase in \ab{the maximum field error, measured as} $\max(\mathbf{B}\cdot \mathbf{\hat{n}}/ B)$, and in the maximum strain in the HTS tape stack, shown in Figure~\ref{fig:finite-build-strain}. A major issue is that the strain in some filaments is above the strain threshold, which would degrade the HTS tape critical current, \ab{\textit{i.e.}} the amount of current the HTS tape can carry before losing superconductivity.\citep{nickel_2021}} 

To remedy this, we use the new finite-build modeling capabilities to refine CSX SFC's. We found that re-optimizing only the coil orientation or geometry was not sufficient to reduce the field error of finite-build coils on the plasma boundary. Instead, \ab{the Boozer surface approach is again used to} refine each configuration. \av{Initial guesses for the optimizations are the SFC obtained as an output of the optimization loop described in Sec.~\ref{sec.optimization}.} The same weights are used when optimizing with a finite-build coil model, but now the optimization takes into account finite-build effects on the magnetic field and HTS strain. Further refinements to the weights could be made on the most promising MFC to try and improve the results of the finite-build refinement. 

The QS error, field error, and HTS strain are presented in Table~\ref{tab:csx_model_comparison} for the best configuration. \ab{Here, the QS error is given by
\begin{equation}
    f_{QS} = \left( \sum_{m,n\neq 0} B_{mn}^2 \middle/ \sum_{m,n}B^2_{mn}\right)^{1/2},
\end{equation}
where $B_{mn}$ is the Fourier mode of the magnetic field strength evaluated on the plasma boundary.} The parameters are expressed across 3 different models: (1) filamentary coils, (2) finite-build coils before refinement, and (3) finite-build coils, after refinement. This demonstrates that we can mitigate finite-build coil effects and reach levels of performance similar to filamentary coils if finite-build effects are considered during optimization.

\begin{table*}
    \begin{tabular}{||c|c|c|c||}
        \toprule 
        Parameter & \multicolumn{3}{c||}{Model type} \\
        \hline
        ~ & Single filament & Finite-build & Finite-build re-optimized \\
        \hline      
        $f_{QS}$ & 0.0785 & N/A &  0.077 \\
        Max $\mathbf{B}\cdot\mathbf{\hat{n}}/B$ error & 0.0031 &  0.0077 & 0.0029 \\
        Max strain $(\%)$ & 0.22 & 0.62 & 0.22 \\
        
        \hline 
        \hline
    
    \end{tabular}
    \caption{Main metrics for CSX selected configuration compared across coil models. The QS metric is undefined for the finite-build configuration before refinement as the plasma boundary is not consistent with the finite build coils.}
    \label{tab:csx_model_comparison}
\end{table*}

\section{The CSX configuration}
\label{sec.final_configuration}

A total of 19 different configurations for CSX were generated. For each configuration, their rotational transform, enclosed plasma volume, coil-to-coil distance, coil-to-plasma distance, coil-to-vessel distance, maximum HTS strain, and maximum force on the HTS tape were evaluated. Based on this analysis, a single configuration was ultimately selected as the most balanced compromise between engineering feasibility and physics objectives. This configuration forms the basis of the present CSX design.

\begin{figure}
    \centering
    \includegraphics[width=\linewidth]{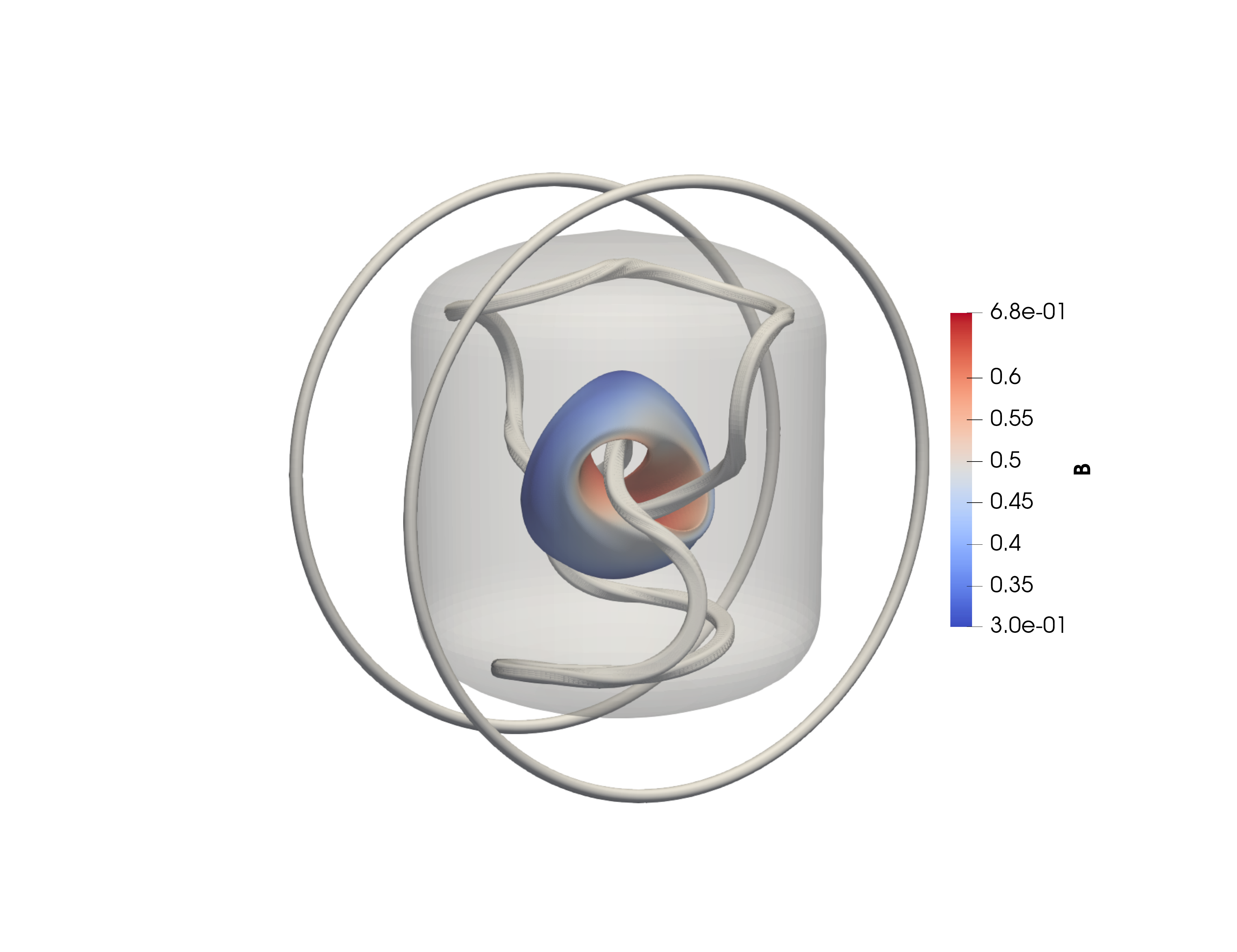}
    \caption{Three-dimensional rendering of the final CSX configuration. Colors on the plasma boundary indicate the magnetic field strength, highlighting the non-axisymmetric structure and quasi-symmetry properties of the configuration.}
    \label{fig.csx_3d}
\end{figure}

A three-dimensional rendering of the final configuration is shown in Figure~\ref{fig.csx_3d}. The plasma boundary is colored according to the magnetic field strength, which was scaled to reach a target magnetic field of $0.5$T on the magnetic axis.

The principal geometrical and engineering characteristics of the configuration are summarized in Table~\ref{tab.csx_figures_of_merit}. The device features a major radius of 0.253\,m and a minor radius of 0.139\,m, corresponding to an aspect ratio of 1.82 and a plasma volume of 0.097\,m$^3$. The coil system consists of two field periods and a minimal coil-to-plasma separation of 0.10\,m, sufficient to accommodate thermal shielding and a vacuum vessel conformal to the coils. The maximum force on the coils is below 2.2\,kN, well within the limits for mechanical support. Coil lengths and currents were selected to remain compatible with practical high-temperature superconducting (HTS) winding and current-carrying capabilities.

\begin{table*}
    \centering
    \begin{tabular}{||ll||ll||}
        \toprule
        Major radius            & 0.253\,m    &   Coil length                     & 5.0\,m         \\
        Minor radius            & 0.139\,m &   IL coil current                 & 316\,kA           \\
        Aspect ratio            & 1.82   &   PF coil current                 & 135\,kA        \\
        Plasma volume           & 0.097\,m$^3$ &   Number of HTS winds per coil & 1000     \\
        Number of field periods & 2     &   Number of HTS stacks         & 4         \\
        Coil-to-coil distance   & 0.287\,m & Coil-to-plasma distance & 0.10\,m \\
        Coil-to-vessel distance & 0.102\,m &  Max force on coils      & 2.18\,kN   \\
        \hline
        \hline
    \end{tabular}
    \caption{Main CSX metrics and engineering parameters. Geometric dimensions and coil characteristics reflect the final optimized configuration.}
    \label{tab.csx_figures_of_merit}
\end{table*}

\begin{figure}[htbp]
    \centering
    \includegraphics[width=\linewidth]{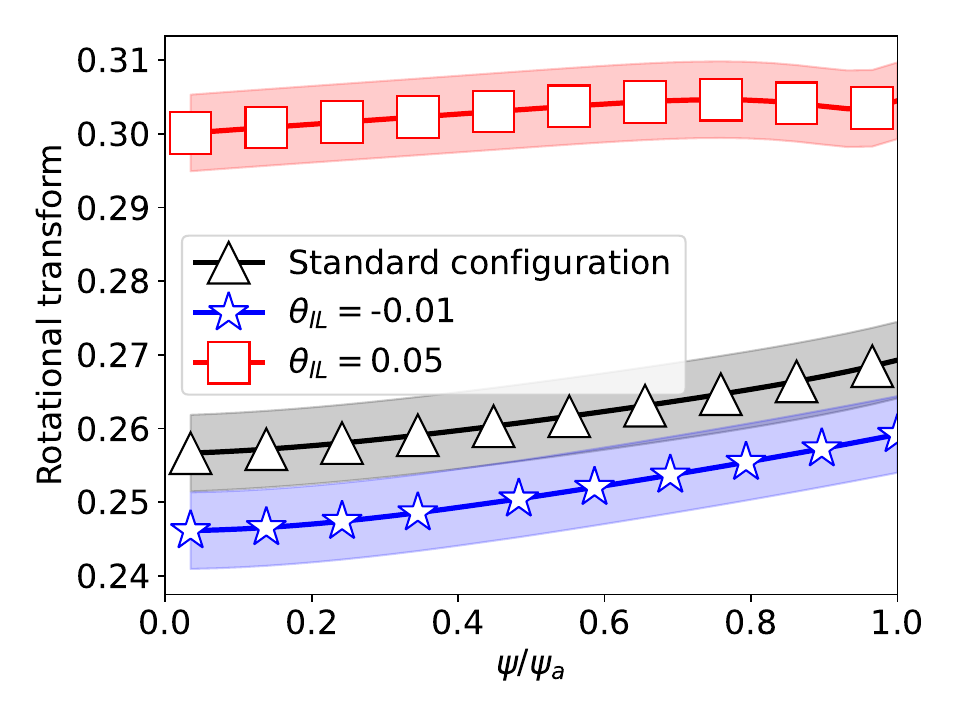}\\
    \includegraphics[width=\linewidth]{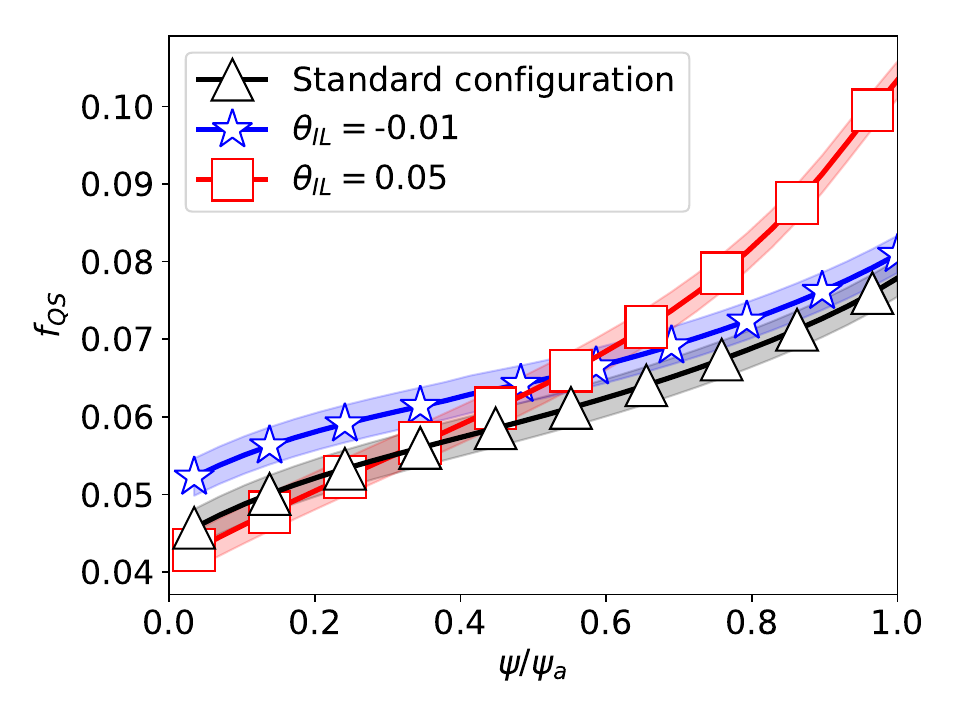}
    \caption{Rotational transform (top) and quasi-symmetry error (bottom) profiles for the final CSX configuration. Different colors correspond to configurations obtained by rotating the IL coils. Shaded regions represent estimated uncertainties arising from manufacturing and alignment errors.}
    \label{fig.vacuum_profiles}
\end{figure}

Magnetic field properties of the selected configuration are shown in Figure~\ref{fig.vacuum_profiles}. The top panel displays the rotational transform profile $\iota(s)$, while the bottom panel shows the QA error. Both quantities are plotted as a function of normalized toroidal flux, with shaded regions indicating expected variations due to manufacturing and assembly tolerances --- see section \ref{sec.sensitivity}. The different colored curves correspond to configurations obtained by rotating the IL coils around their axes, as we will discuss in section \ref{sec.flexibility}. The profiles indicate that the configuration is free of low-order resonances in the confinement region and maintains a relatively small QS error throughout the plasma volume, below the target value of $8\%$ discussed in \citet{baillod_integrating_2025}.

\begin{figure}[htbp]
    \centering
    \includegraphics[width=\linewidth]{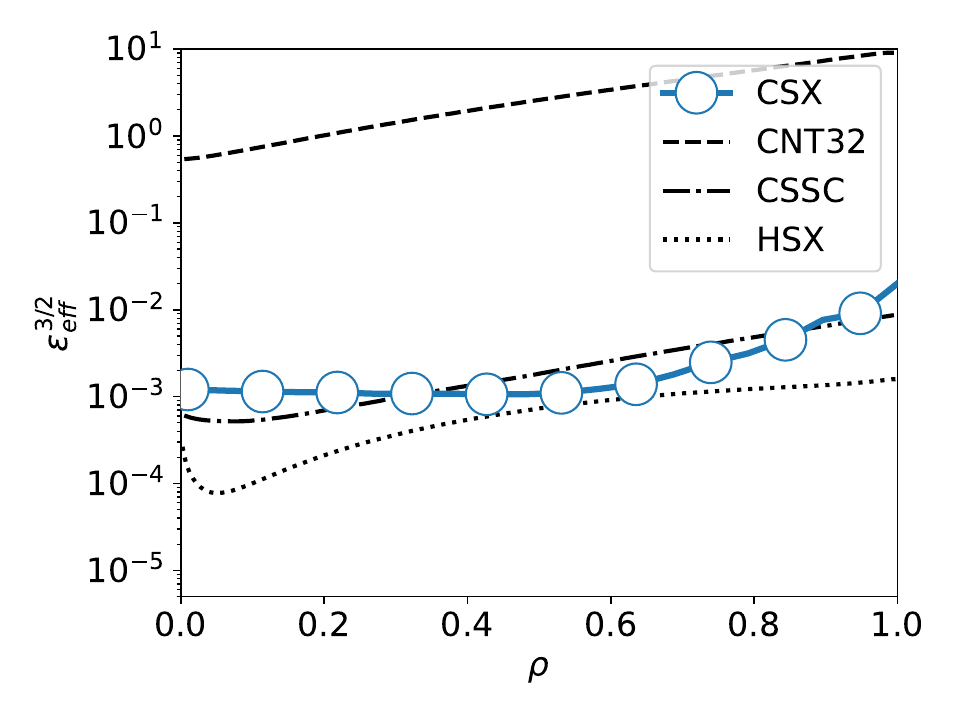}
    \caption{Comparison of the effective ripple, $\epsilon_{\mathrm{eff}}$, for CSX and other compact stellarators.}
    \label{fig.epsilon_effective}
\end{figure}

A common metric used to characterize neoclassical transport in stellarators is the effective ripple, $\epsilon_{\mathrm{eff}}$, defined by Nemov \textit{et al.}\citep{nemov_1999a}. Figure~\ref{fig.epsilon_effective} compares the effective ripple of CSX with that of several compact stellarators of similar size, including CNT, CSSC, and HSX. The CSX configuration achieves a substantial reduction in $\epsilon_{\mathrm{eff}}$ relative to CNT, indicating improved neoclassical confinement. The obtained values are comparable to those achieved in devices of similar dimension such as CSSC and HSX. This result demonstrates the effectiveness of the optimization strategy employed in achieving favorable confinement properties while satisfying engineering constraints.

\section{Sensitivity studies} \label{sec.sensitivity}


Real coils inevitably deviate from their design specifications due to manufacturing imperfections and \ab{misalignment during installation}. To assess the robustness of CSX, we perform a sensitivity analysis examining three physically distinct error sources, applied independently to each coil, \textit{i.e.} allowing for breaking of stellarator symmetry.

Geometric errors along the coil path are modeled as the superposition of two Gaussian processes (GPs) with squared exponential kernels,~\citep{wechsung_2022} capturing deviations at different spatial scales. CSX coils will be 3D printed in an Aluminium alloy in 20 different pieces of around $25$\,cm each. A first set of GP is used to model 3D-printing error with corresponding amplitude $\sigma_{\text{mfg}} = 0.3$\,mm and correlation length $\ell_{\text{mfg}} = 25$\,cm. Long-correlation perturbations capture smooth, global deviations from conductor placement within the finite-build winding pack, with amplitude $\sigma_{\text{wind}} = 0.5$\,mm and correlation length $\ell_{\text{wind}} = 5$\,m. The sum of these two GPs approximates the full spectrum of manufacturing errors expected in the CSX coil fabrication.

Installation errors produced during the assembly CSX are modeled with rigid-body translations and rotations. Translations amplitude are sampled from a Gaussian distribution with standard deviation $\sigma_{\text{spatial}}$, while rotations are sampled with standard deviation $\sigma_{\text{angle}}$. \ab{Both the rotation axes and the translation direction are drawn uniformly on the unit sphere.} The full set of errors used to model manufacturing and installation errors is summarized in Table \ref{tab:csx_errors}.

\begin{table*}
    \begin{tabular}{||c|c|c||}
        \toprule
        Error Type & Parameters  & Model \\
        \hline
        Installation & $\sigma_\text{spatial} = 1$ mm, $\sigma_\text{angle} = 1.0°$  & Rigid body \\
        Manufacturing & $\sigma_{mfg} = 0.3$mm, $l_{mfg}=28$cm & GP, localized \\
        Winding & $\sigma_\text{wind} = 0.5$mm, $l_\text{wind}=5$m & GP, full-coil \\
        \hline
        \hline
    \end{tabular}
    \caption{Error specifications for CSX coil tolerances.}
    \label{tab:csx_errors}
\end{table*}

To establish coil installation tolerances, \textit{i.e.} the maximum $\sigma_\text{angle}$ and $\sigma_\text{spatial}$ CSX can tolerate, we conduct a Monte Carlo sensitivity analysis by generating an ensemble of randomly perturbed coil configurations. \ab{Each perturbed coil set was generated by applying the two GP processes, the rigid body rotation, and the rigid body translation.} For each point in the $(\sigma_\text{spatial},\sigma_\text{angle})$ parameter space, we generate $N$ independent realizations, yielding a statistical ensemble from which we extract \ab{the distribution of quasi-symmetry error and the edge rotational transform.}

\subsection{Success Criteria}

\begin{figure}
    \centering
    \includegraphics[width=\linewidth]{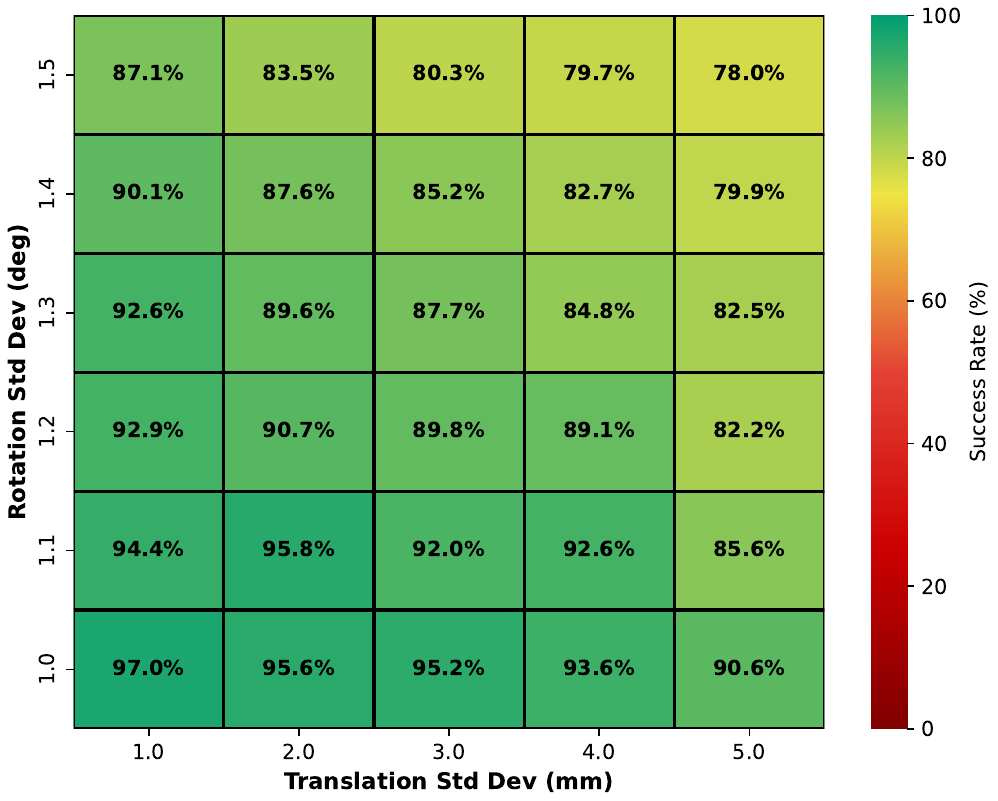}
    \caption{Success rate, \textit{i.e.} proportion of perturbed configuration that fit the design specifications.}
    \label{fig.montecarloscatter}
\end{figure}

We define a configuration as successful if it satisfies $f_\text{QS} < 0.08$, and $0.255 < \iota_\text{edge} < 0.33$. By constraining $0.255 < \iota_{\text{edge}} < 0.33$, the rotational transform profile avoids the $\iota = 1/4$ and $\iota = 1/3$ rational surfaces throughout the plasma volume. The tolerance is identified as the largest set of perturbation amplitude $(\sigma_\text{spatial},\sigma_\text{angle})$ for which the success probability $P_\text{success} = N_\text{success}/N$ exceeds 90\%, given that the Boozer surface is successfully solved. This analysis yields possible tolerances of $\sigma_\text{angle} = 1.1\degree$ and $\sigma_\text{spatial} = 3$~mm (Fig.~\ref{fig.montecarloscatter}), establishing the constraints that must be satisfied during assembly. 

\begin{figure}
    \centering
    \includegraphics[width=\linewidth]{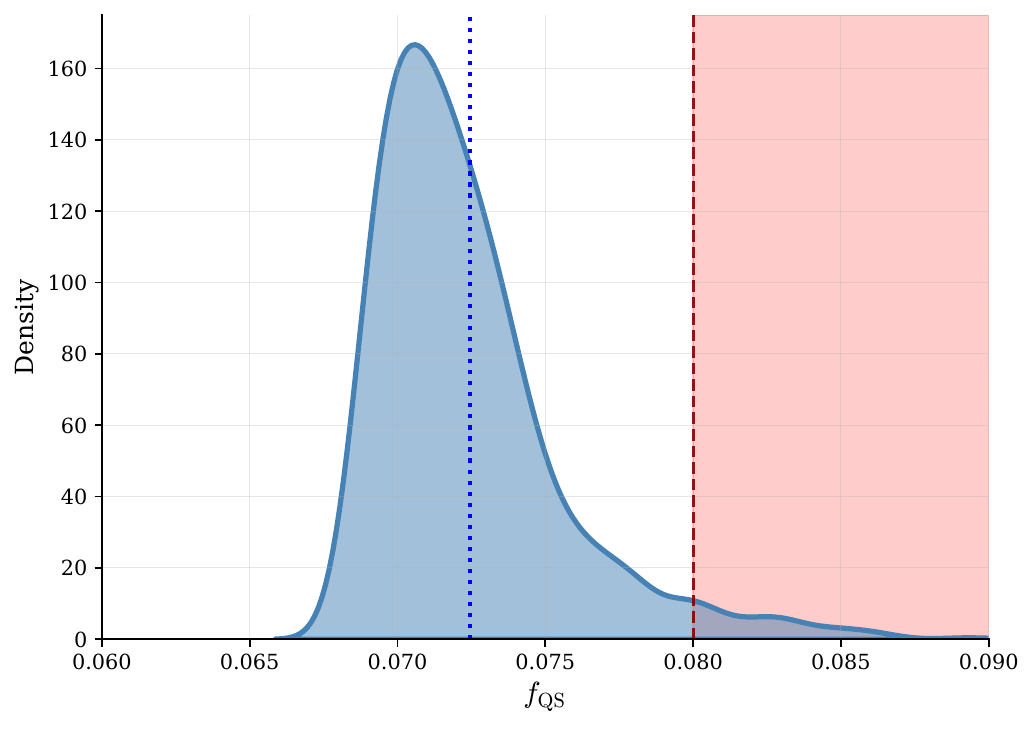}
    \includegraphics[width=\linewidth]{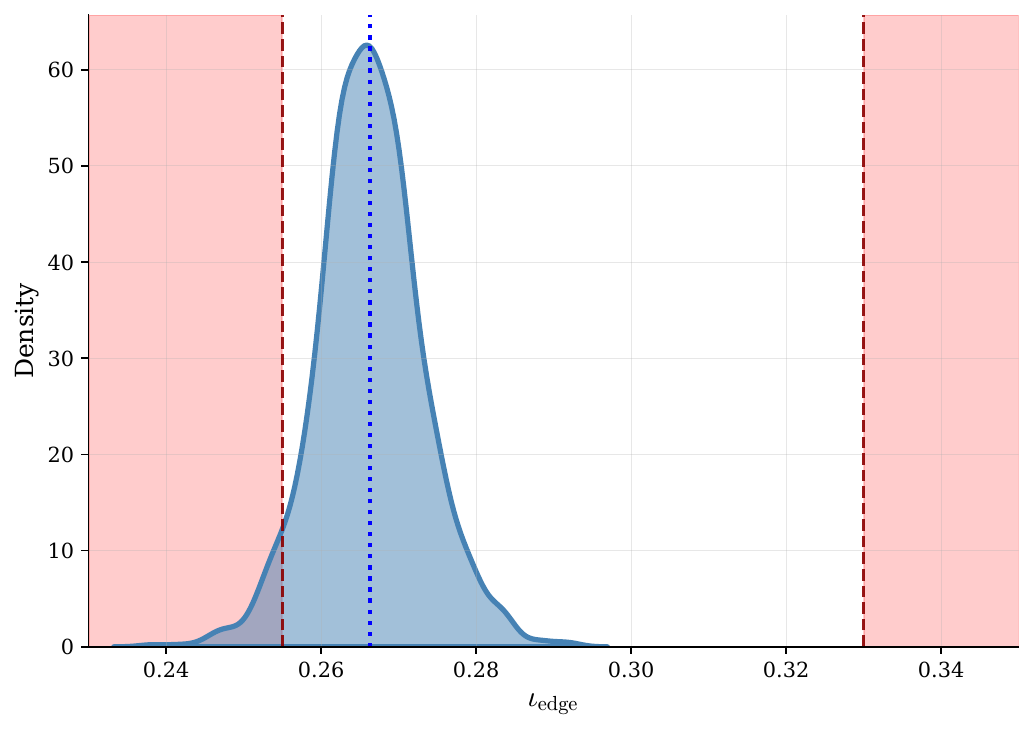}
    \caption{Distribution of $f_{QS}$ (top) and $\iota_\text{edge}$ (bottom) for $\sigma_{spatial}= 3mm$ and $\sigma_{angle} =1.1 \degree$ perturbation. The blue dotted line represent the unperturbed configuration.}
    \label{fig.error3mm1deg}
\end{figure}

The distribution of QS error and edge rotational transform for $(\sigma_\text{spatial},\sigma_\text{angle})=(1.1\degree,3\text{mm})$ are shown on Figure \ref{fig.error3mm1deg}. As expected, only a small fraction of perturbations break the conditions for success. Interestingly, there are some perturbations that decrease the QS error, indicating that there are configurations with lower QS error that the optimizer could not reach, either because of engineering constraints, numerical resolution, or enforcing stellarator symmetry. Note that coils perturbations that increase the rotational transform are unlikely --- the condition $\iota<0.33$ is never broken.

As a complementary approach, the shape gradient\citep{landreman_computing_2018} of some critical metrics is evaluated as a linear approach to determining the tolerances of the coils.
The shape gradient quantifies how sensitive a figure of merit is to infinitesimal perturbations of the coil geometry, providing a spatially resolved map of where deformations most strongly affect plasma performance. Shape gradients identify critically sensitive coil regions where small geometric changes disproportionately degrade the magnetic field, directly informing where to concentrate manufacturing precision and measurement efforts.

\begin{figure*}
    \centering
    \includegraphics[width=\linewidth]{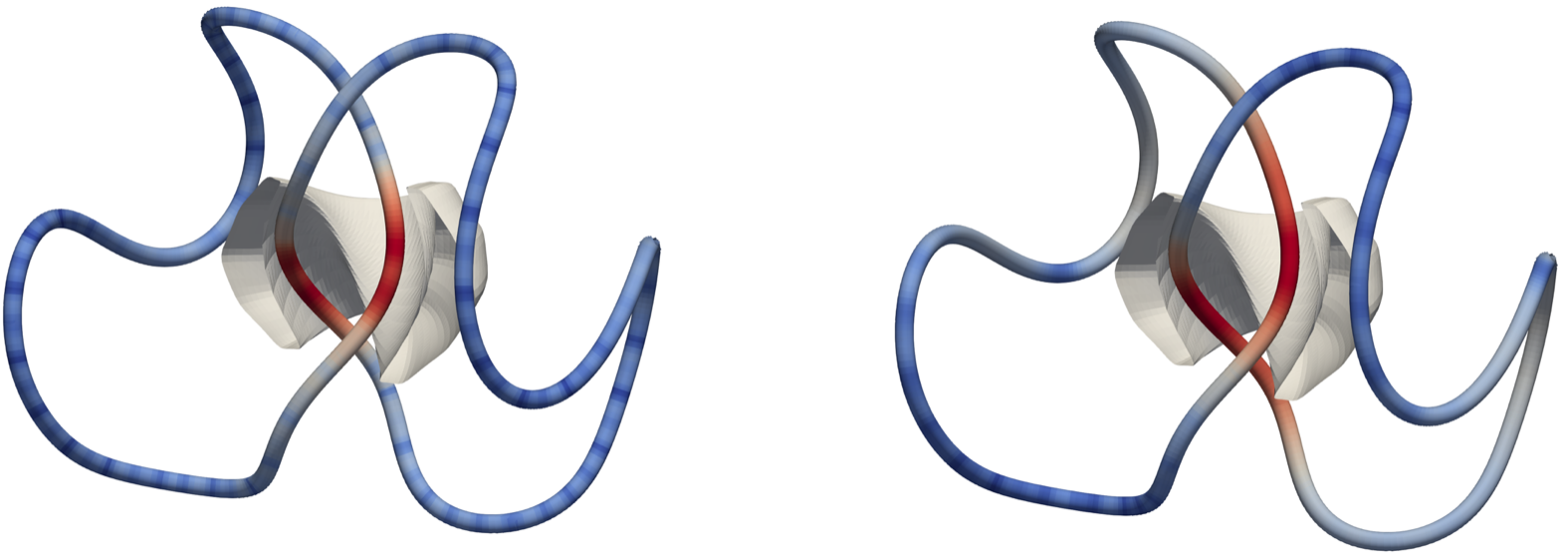}
    \caption{Shape gradient sensitivity visualization for (left) edge rotational transform $\iota_{\text{edge}}$ and (right) quasi-symmetry error $f_{\text{QS}}$. Color intensity indicates the local sensitivity magnitude along the coil path, identifying regions where geometric perturbations most strongly affect each metric. High-sensitivity regions (warm colors) require tighter manufacturing tolerances, while low-sensitivity regions (cool colors) are more forgiving of geometric deviations.}
    \label{fig:shape_gradient_plot}
\end{figure*}

The shape gradient is evaluated following the work of \citet{landreman_computing_2018}. Shaped gradients related to the edge rotational transform and the QA error are shown on Figure \ref{fig:shape_gradient_plot}. From these gradients, we compute a uniformly distributed tolerance, the maximum perturbation magnitude that can be applied equally along the entire coil length while keeping the figure of merit within acceptable bounds. The uniformly distributed tolerances are $7.778$~mm for QS error and $2.324$~mm for $\iota_{\text{edge}}$, indicating that rotational transform is approximately three times more sensitive to coil deformation. This hierarchy is consistent with the Monte Carlo results, where the $\iota_{\text{edge}}$ constraint yields the lowest pass rate. The shape gradient tolerances are somewhat tighter than the Monte Carlo-derived values, as expected as the shape gradient assumes the worst possible coil deformation.    




\ab{Sensitivity studies show that CSX has relatively tight tolerances on coil positioning, in the range of what can be achieved for this kind of experiment. It is important to note that even if these tolerances are not met, error field correction coils could be designed \textit{a posteriori} to fix large errors, and better control the rotational transform profile.}


    

\subsection{Achieving experimental flexibility}\label{sec.flexibility}
Experimental flexibility in the plasma geometry is essential for interpreting measurements across multiple configurations. In CSX, such flexibility can be introduced through two degrees of freedom: varying the current ratio between the IL and PF coils, and rotating the IL coil set. These approaches mirror the strategy originally proposed for obtaining difference configurations from the CNT coils.\citep{kremer_2003}

For the CSX design, we base the experimental flexibility on the rotation of the IL coils, parameterized by an angle $\theta_{IL}$. The reference configuration corresponds to $\theta_{IL}=0$, i.e., the optimized CSX design without coil rotation. Finite values of $\theta_{IL}$ impose a rigid-body rotation of the IL coil set about the symmetry axis of the vacuum vessel. Decreasing this angle brings the coils progressively closer to the plasma boundary; beyond a certain threshold, the coils intersect the last closed flux surface (LCFS) and act as a limiter, thereby reducing the enclosed plasma volume.

Current designs allow for continuous rotation of the IL coils by mounting the  coils on rotatable flanges. This enables an entire coil assembly and corresponding cooling system (as described in Schmeling \textit{et al.}\citep{schmeling_2025}) to be rotated around the axis relative to the other, stationary coil. To ensure precise alignment at chosen angles, such as the reference $\theta_{IL}=0$ and other angles of interest, the coils will  be joined at their crossover section using an indexing pin joint. 

\begin{figure}
\centering
\includegraphics[width=\linewidth]{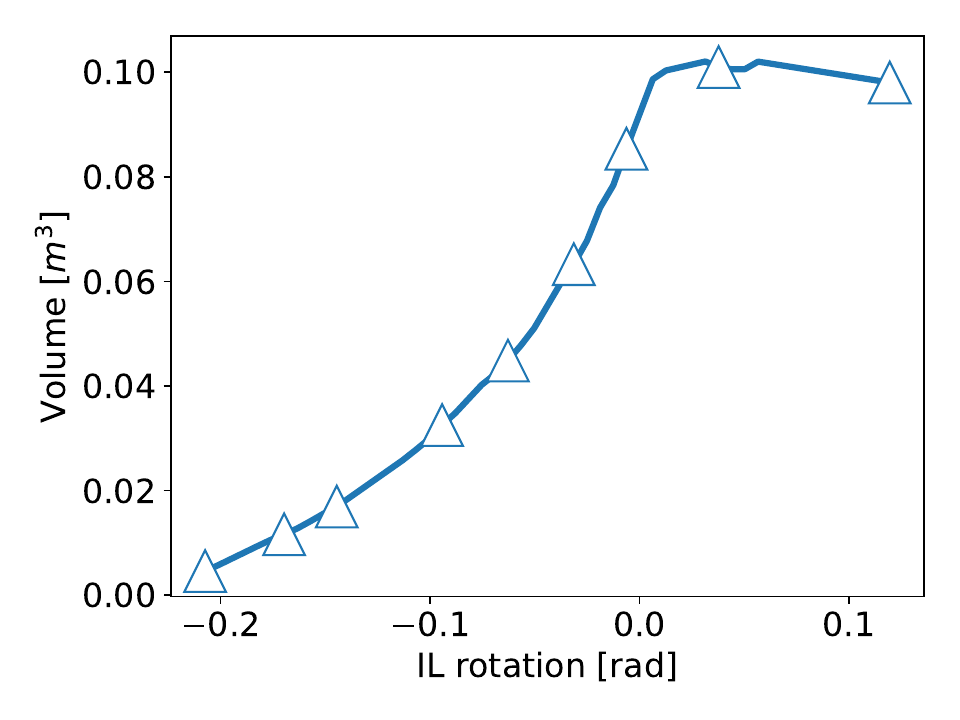}
\caption{Enclosed plasma volume as a function of IL-coil rotation angle $\theta_{IL}$.}
\label{fig:flexibility_volume}
\end{figure}

We now present a scan of the angle $\theta_{IL}$. For each configuration, we identify the LCFS as the first magnetic surface with a minimal coil-to-plasma distance of $10$\,cm. When no intersection occurs, the LCFS is defined as the outermost existing closed surface. The resulting enclosed plasma volume is shown in Figure ~\ref{fig:flexibility_volume}, which demonstrates a systematic reduction as $\theta_{IL}$ decreases.

\begin{figure}[h]
\centering
\includegraphics[width=\linewidth]{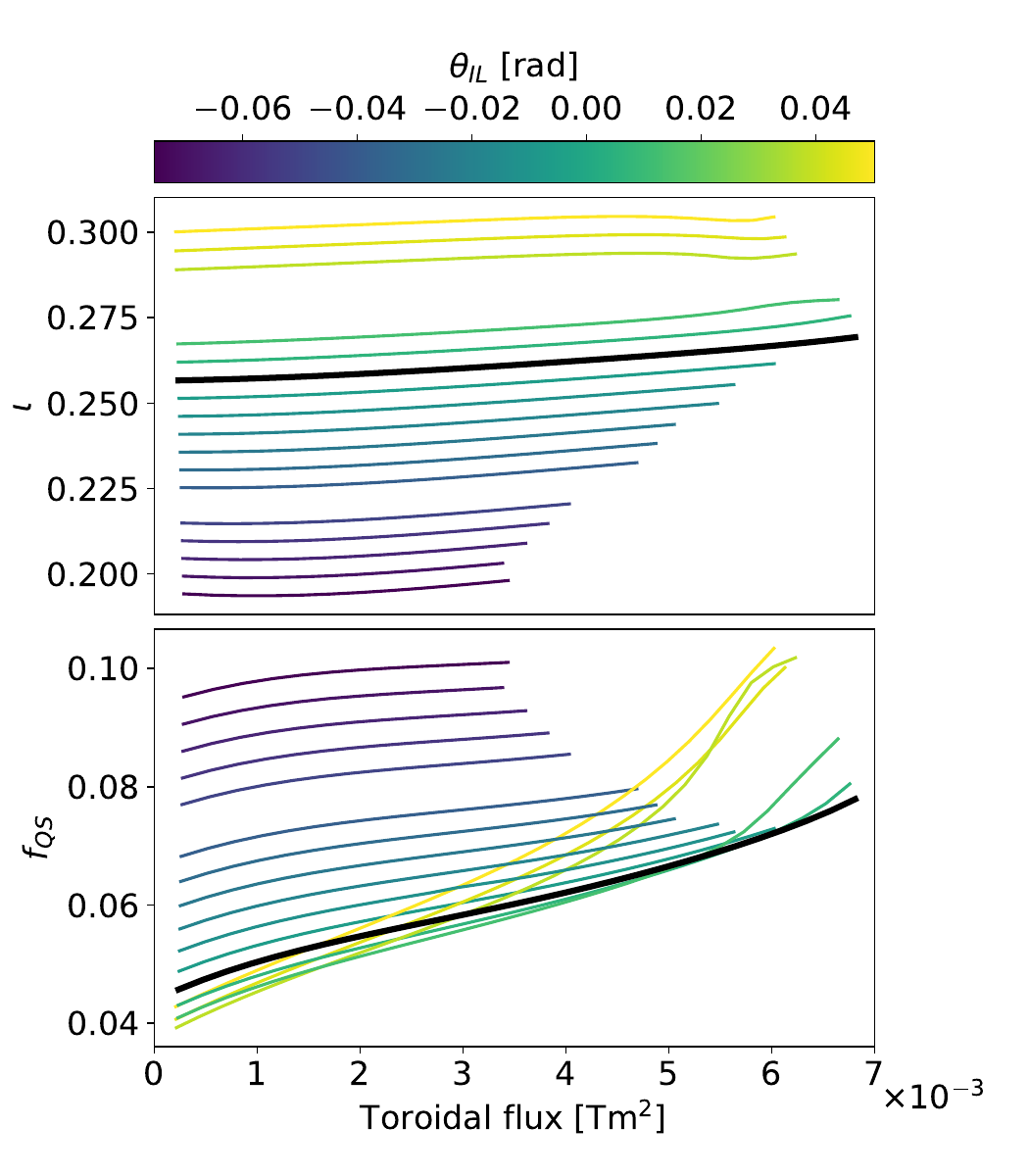}
\caption{Rotational-transform (top) and quasi-symmetry error (bottom) profiles for configurations obtained by rotating the IL coils. }
\label{fig:flexibility_profiles}
\end{figure}

The corresponding rotational-transform and QS error profiles are presented in Fig. \ref{fig:flexibility_profiles}. Only configurations with a non-negligible plasma volume ($V>0.05\text{m}^3$) are shown. The scan reveals that a broad range of rotational transform profiles can be accessed through IL-coil rotation. In particular, configurations with low order rational surfaces can be obtained, enabling the emergence of large magnetic islands, which could be of interest for magnetic island physics and error field correction studies.

In addition, configurations with significantly degraded quasi-symmetry can be achieved. This controllable departure from QA is a desirable capability for future CSX experiments, as it enables systematic studies of neoclassical transport, flow damping, and symmetry-breaking effects.

Although the IL coils are designed such that any value of $\theta_{IL}$ can be obtained, we highlight two configurations of interest. The first is obtained by rotating the IL coils by an angle of $\theta_{IL}=-0.013$ radians. This configuration has a rotational transform profile that crosses the resonance $(n,m)=(1,4)$, thereby featuring large magnetic islands in the plasma core. Another configuration of interest can be obtained by rotating the coils in the opposite direction, to an angle of $\theta_{IL}=0.05$ radians. In this case, the rotational transform is larger ($\iota\approx0.301$), and the QS error is larger at the plasma edge, $f_{QS}\approx 9.5\%$. This could be a useful configuration to study effect of QS-breaking on neoclassical flow damping measurements.

\section{Neoclassical calculations with SFINCS}

 As part of the goal of CSX to validate predicted properties of stellarators close to quasi-axisymmetry, an intended experiment is to measure  the rate of flow damping along the toroidal direction; it is expected for quasi-symmetric stellarators that the neoclassical flow damping along the direction of symmetry is reduced relative to unoptimized stellarators.\citep{helander_2008} The aim of the neoclassical flow modeling is to predict a time-scale over which a flow induced in CSX may be damped, and so verify that such flow damping will be measurable in CSX. A secondary aim is to confirm that the neoclassical heat flux predicted for CSX is less than the heat flux predicted for CNT.

The intended experiment to measure flow damping follows a procedure similar to the one proposed by \citet{gerhardt_2005}. \ab{At first, the plasma is unbiased, and the electric field is given by the ambipolarity condition. 
The electrostatic potential on a magnetic surface can then be biased by inserting in the probe in the plasma.}
The shift from an ambipolar state to a state with nonzero radial current generates a $\vv J \times \vv B$ force, which in turn supplies a torque. This causes some shift in the flow along the direction of symmetry. If the biasing potential is then removed, the flow will be damped until the flow returns to the original steady state. The flow is measured throughout this process by a Mach probe or similar diagnostic. A flow damping rate can then be measured provided it is slow relative to the diagnostic sampling rate. We show below that this should be the case for CSX.

\begin{figure}
    \centering
    \includegraphics[width=0.9\linewidth]{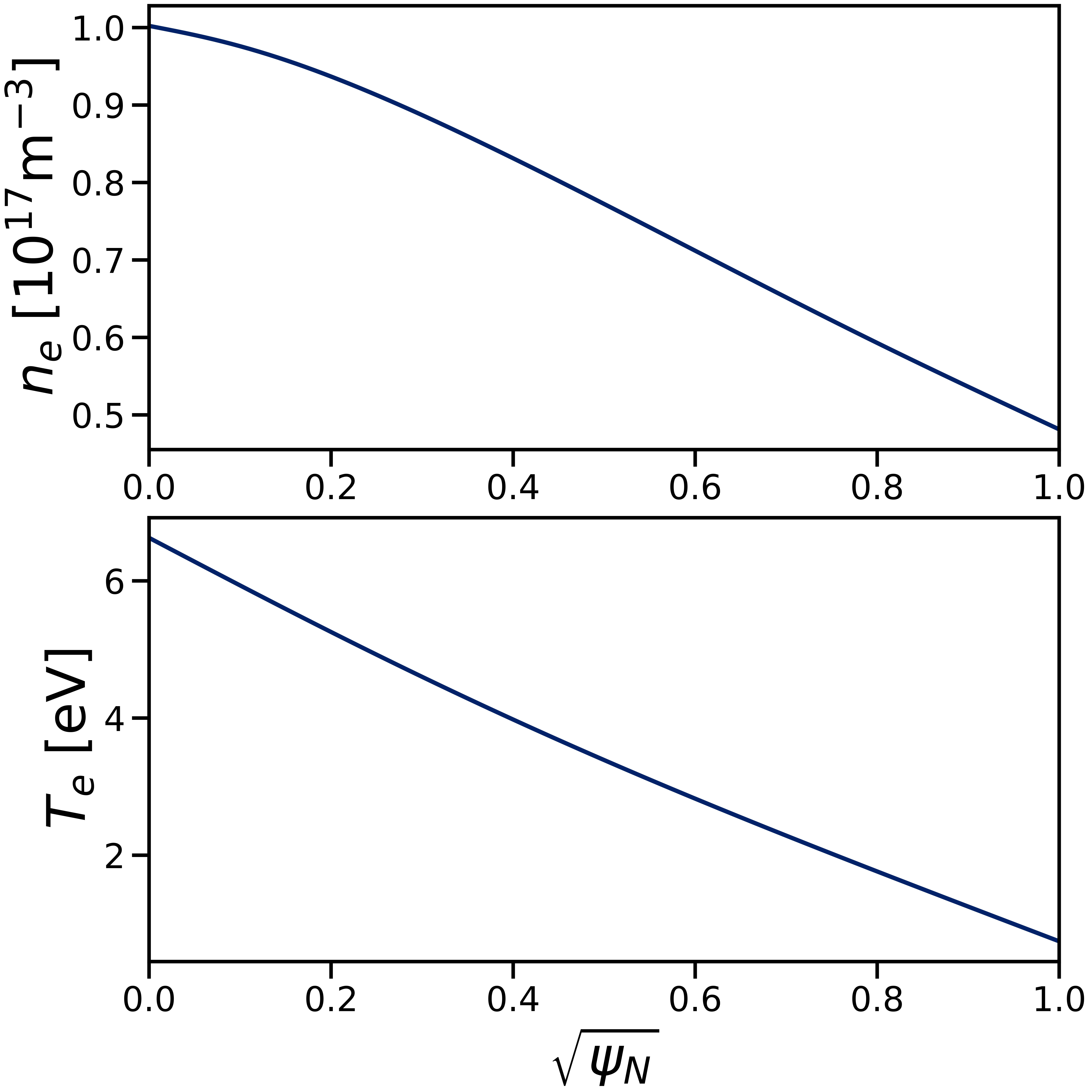}
    \caption{Electron density and temperature profiles used for SFINCS calculations. The ion density is equal to the electron density and the ion temperature is set to 30\% the electron temperature.}
    \label{fig:sfincs_profiles}
\end{figure}

\begin{figure}
    \centering
    \includegraphics[width=1.0\linewidth]{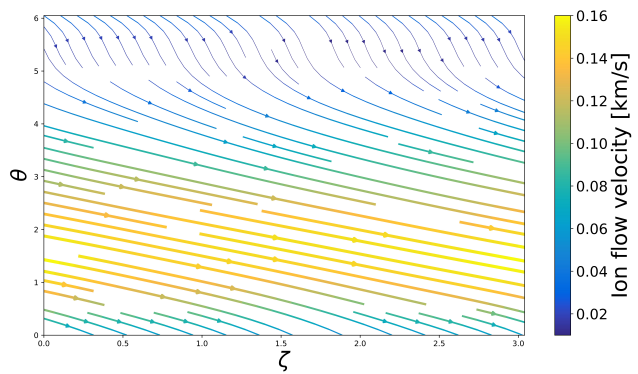}
    \caption{Flow velocity $v_{||}\vv{\hat b} + \vv{v_\perp}$ in CSX on $\sqrt{\psi_N} = 0.55$ with $v_{||}$ calculated with SFINCS and $\vv v_\perp$ from Eq. \ref{eqn:vperp}. Here, $\theta$ and $\zeta$ are the poloidal and toroidal Boozer angles. The visualization assumes the covariant basis is orthonormal and aligned with the axes.}
    \label{fig:flow}
\end{figure}

To model the experiment, we perform neoclassical calculations using the Stellarator Fokker-Planck Iterative Neoclassical Conservative Solver (SFINCS).\citep{landreman_2014} SFINCS solves the drift-kinetic equation given density, temperature, and radial electric field $E_r$ profiles, in addition to the CSX equilibrium. Figure \ref{fig:sfincs_profiles} shows the density and temperature profiles used for this model, chosen to agree with the expected plasma parameters for CSX. \cite{baillod_integrating_2025} We assume a deuterium plasma and do not account for impurities. The ambipolar $E_r$ is estimated by running SFINCS for a range of $E_r$ values and choosing the value such that the radial current is smallest. Useful quantities obtained from SFINCS include the parallel flow velocity $v_{||}$, the radial neoclassical heat flux $Q_N$, and the quantity $\fsa{\vv J \cdot \nabla \psi}$. We obtain the vector flow velocity as 
\begin{equation}
    \vv v = v_{||}\vv{\hat b} + \vv v_{\perp},
\end{equation} where 
\begin{equation}\label{eqn:vperp}
    \vv{v}_{\perp } = \left(\frac{\partial\Phi}{\partial\psi} + \frac{1}{Z_in_i}\frac{\partial p_i}{\partial\psi} \right)\frac{\vv{B}\times\nabla\psi}{B^2}
\end{equation}
is the cross-field flow term obtained from frictionless momentum balance with isotropic pressure, $\Phi$ is the plasma potential, and $Z_i$, $n_i$, and $p_i$ are the ion charge, density, and pressure respectively. \citep{helander_2002a} The flow on the $\sqrt{\psi_N}=0.55$ surface is plotted in Figure \ref{fig:flow}. The main component of the flow is in the toroidal direction, as expected for a stellarator close to QA. Note that only the Pfirsch-Schluter component of the flow is expected to be in the symmetry direction.\citep{Landreman2012} 

Suppose that the plasma has been "spun up" to steady state using an electric field $E_r$ generated by the biasing probe. The flux-surface-averaged component of the torque density relevant to toroidal rotation is 
\begin{equation}\label{eqn:ntv}
    \fsa{\tau} =\fsa{\vv J \times \vv B\cdot \pdv{\vv r}{\zeta}} = \iota \fsa{\vv J \cdot \nabla \psi},
\end{equation}
with $\zeta$ the toroidal Boozer angle. Since we know the plasma flow velocity $\vv v$ everywhere on a flux surface, we determine the flux-surface-averaged toroidal angular momentum density 
\begin{equation}\label{eqn:amd}
    \fsa{l} = \fsa{m_in_i\vv v\cdot \pdv{\vv r}{\zeta}},
\end{equation}
where $m_i$ is the ion mass. Because $\fsa{\tau}=\frac{d\fsa{l}}{dt}$, we estimate the flow damping time scale as
\begin{equation}\label{eqn:tfd}
    t_{fd}(E_r) = \left | \frac{\langle l \rangle (E_{r}) - \langle l \rangle(E_{r0})}{\frac{1}{E_{r0}-E_r}\int_{E_{r0}}^{E_{r}}\langle \tau \rangle(E_r^\prime)dE_r^\prime} \right |,
\end{equation}
where $E_{r0}$ is the ambipolar $E_r$. Note that since we have an expression for $\vv v$, then $\fsa{l}(E_r^\prime)$ is obtained directly through Eq. \ref{eqn:amd} with $\vv v$ the velocity obtained when running SFINCS with $E_r = E_r^\prime$. This $t_{fd}$ is the time it would take for the plasma to rotate from its "spin up" state (with radial electric field $E_r^\prime$) back down to its ambipolar state (with radial electric field $E_{r0}$).

From this modeling, the flow damping time scale for CSX is estimated to be on the order of tens of milliseconds. Specifically, for a probe biasing scenario measuring flow damping near $\sqrt{\psi_N} = 3/4$, a "spin up" state with $E_r \approx -60$\,V/m has an estimated flow damping time of about 40\,ms. The ambipolar $E_r$ on $\sqrt{\psi_N} =3/4$ is approximately $-28$\,V/m. The sampling rate of plasma diagnostics available to measure flow is often in the kHz to MHz range; such resolution is more than sufficient to experimentally observe the predicted flow damping time.

To validate the results of this analysis, we applied the same methods to HSX and showed reasonable (order of magnitude) agreement with the neoclassical predictions made by \citet{gerhardt_2005}, where the predicted flow damping time in HSX exceeded the measured time by an order of magnitude. It might then be expected that our prediction for CSX is similarly an overestimate -- if this is measured to be the case, it would suggest that the flow damping is not set purely by neoclassical physics, and that additional interactions (e.g. collisions with neutrals) contribute significantly. Note that an overestimate of this magnitude would still place the flow damping time well within the measurable range. Additionally, although the reproduction of the modeling results from HSX is promising, it is no guarantee the \textit{ad hoc} flow damping time estimate in Eq. \ref{eqn:tfd} is accurate.  

In addition, neoclassical calculations are performed with the same profiles (adjusted to the ambipolar $E_r$) for CNT. Comparison of the total heat loss through the LCFS from CSX, $Q_N^{CSX}$, to the heat loss in CNT, $Q_N^{CNT}$, shows $Q_N^{CSX}/Q_N^{CNT} \approx 0.16$, confirming one should expect reduced neoclassical heat flux in CSX as compared to its unoptimized predecessor. Estimating the turbulent heat loss $Q_T$ in CSX using a gyro-Bohm estimate gives $Q_N^{CSX}/Q_T \approx 0.8$, additionally indicating neoclassical heat flux in CSX may be on the order of the turbulent heat flux. Note that the gyro-Bohm scaling is an estimate for microturbulence heat flux; it does not take into account macro-turbulence that could appear in CSX due to the lack of a magnetic well.

Further work, including a sensitivity analysis with respect to profiles and more sophisticated modeling, would be necessary to make a robust prediction of the flow damping time. The analysis presented above is however sufficient to confirm an order-of-magnitude estimate and thus experimental feasibility. 

\section{Conclusion}
The CSX configuration presented above meets the primary design objectives that were set at the outset. The resulting configuration is close to quasiaxisymmetry and it is therefore expected that a reduction in neoclassical flow damping along the symmetry direction will be experimentally measurable, as indicated by SFINCS calculations. In addition, experimental flexibility is enabled through rotation of the interlinked (IL) coils, providing valuable experimental control parameters. Finally, the configuration satisfies the relevant engineering constraints, in particular those associated with strain limits on the HTS tape, and has been shown to be sufficiently robust against anticipated manufacturing tolerances and coil positioning errors.

With the CSX design now selected, future work will focus on further prototyping of small-scale coils and, ultimately, on construction of the device. The design and integration of auxiliary systems, including diagnostics, will be addressed in subsequent phases of the project.

\section*{Acknowledgments}
The authors acknowledge useful discussions with Andrew Giuliani, Samuel Lazerson, Dario Panici, the CSX engineering team, and the EPOS team. The authors acknowledge funding from the Simons Foundation Targeted MPS Program (Award 1151685), the PPPL LDRD program, and the Simons Foundation MPS Collaboration Program (Award 560651). This research used resources of the National Energy Research Scientific Computing Center (NERSC), a Department of Energy Office of Science User Facility using NERSC award FES-ERCAP30322.

\section*{Conflict of interest}
The authors report no conflict of interest.

\section*{Data availability}
All data and scripts relevant to this publication are available on the Zenodo archive \textit{10.5281/zenodo.18115389}.

\section*{Bibliography}
\bibliography{csx.bib}

\end{document}